\documentclass[aps,nofootinbib,amsmath,amsfonts,preprint]{revtex4}
\usepackage{hyperref}
\usepackage{graphicx,xcolor}
\usepackage{longtable}
\usepackage{subfigure}
\usepackage{xfrac}

\newcommand{\eqr}[1]{Eq.~\eqref{#1}}

\newcommand{\figr}[1]{Fig.~\ref{#1}}

\newcommand{\velocity}{{\rm{v}}}

\newcommand{\welocity}{{\rm{w}}}
\newcommand{\average}[1]{\left\langle{#1}\right\rangle}

\newcommand{\sss}[1]{{\scriptscriptstyle #1}}

\begin{document}

\title{Thermalization of non-stochastic Hamiltonian systems}
\author{K.~S.~Glavatskiy}
\affiliation{Centre for Complex Systems, The University of Sydney, New South Wales 2006, Australia}
\author{V. L. Kulinskii}
\affiliation{Department of Theoretical Physics, Odessa National University, Dvoryanskaya 2, 65026 Odessa, Ukraine}
\date\today

\begin{abstract}
Ability of dynamical systems to relax to equilibrium has been investigated since the invention of statistical mechanics, which establishes the connection between dynamics of many-body Hamiltonian systems and phenomenological thermodynamics. The key link in this connection is stochasticity, which translates the deterministic evolution of a dynamical system to its probabilistic exploration of the state space. To-date research focuses on determining the conditions of stochasticity for particular systems. Here we propose an alternative agenda and prove thermalization for non-stochastic Hamiltonian systems. This shows that thermalization happens in both stochastic and non-stochastic systems, reducing the need to rely on stochasticity in a ``coarse-grained'' analysis. The result is valid for an arbitrary classical Hamiltonian system and does not rely on the thermodynamic limit or the particular form of the interaction potential. It utilizes the property of adiabatic invariance, and reveals a deep relation between the structure of the microscopic Hamiltonian and macroscopic thermodynamics.
\end{abstract}

\maketitle

%\tableofcontents

%\numberwithin{equation}{section}

%\section{Introduction}

%\section*{Main}

Stochasticity plays crucial role in the foundations of statistical mechanics \cite{Moore2015}. Evolution of a physical system consisting of large number of interacting particles is deterministic, yet such system typically reveals thermodynamic behavior. The latter can be derived from statistical mechanics \cite{ll5}, provided the underlying dynamical system is stochastic. Despite a number of rigorous results \cite{Simanyi2004, Sinai1963} in ergodic theory \cite{SinaiErgodic} and stochastic dynamical systems \cite{ZaslavskyChaos}, ergodicity of a general dynamical system still has the notion of a hypothesis, and for the purposes of statistical mechanics the stochasticity of such system is simply postulated. This leaves open the question of the possibility of thermalization of an arbitrary dynamical system.

In this letter we present an alternative approach to address this question. In particular, we prove that stochasticity is not required for a dynamical system to experience thermalization, i.e. to reveal thermodynamic behavior. To do this we consider an opposite of a stochastic dynamical system, a completely integrable system, and prove that it can also be described using the thermodynamic approach. We further relax the integrability condition and show that the same result also holds for a partially integrable system. The results are presented for a classical Hamiltonian system and do not depend on the type of interactions between the particles.

Consider a dynamical system of $N$ particles which has the energy $E$ and performs finite motion within the volume $V$. The system is either completely or partially integrable and is described by a classical Hamiltonian $H$.  Neither $N$ nor $V$ needs to be large, i.e. the system does not need to be in the thermodynamic limit. Below we derive thermodynamics for such system. In particular, we show, that one can identify the deterministic temperature $T$ and the deterministic pressure $P$, which consistently resemble the thermodynamic temperature and pressure respectively. Furthermore, we show that one can also identify the deterministic entropy $S$, so that the evolution of the system as a whole is described by the standard thermodynamic equation 
\begin{equation}\label{eq/00}
dE = T\,dS - P\,dV
\end{equation}
which expresses the combination of the first and the second laws of thermodynamics for equilibrium processes. 

Thermodynamic approach provides a reduced description of the system, and uses just few thermodynamic variables instead of a large number of dynamic variables (which is of the order of the Avogadro's number). This implies some sort of coarse-graining of the dynamical system. In statistical mechanics this is realized by state space averaging of dynamic variables. An alternative approach is to use time averaging of dynamic variables. Typical experimental measurements are not instantaneous: they represent some mean value over the time of measurement, which is typically much larger than a characteristic microscopic time. In this analysis we will use the following time average of a dynamic variable $x$
\begin{equation}\label{eq/04}
\average{x} \equiv \lim_{\tau \to \infty}\, \frac{1}{\tau} \int_0^{\tau} x(t)\,dt
\end{equation}

\paragraph*{Completely integrable system.}

\begin{figure*}[ht]
\centering
{\includegraphics[scale=0.75]{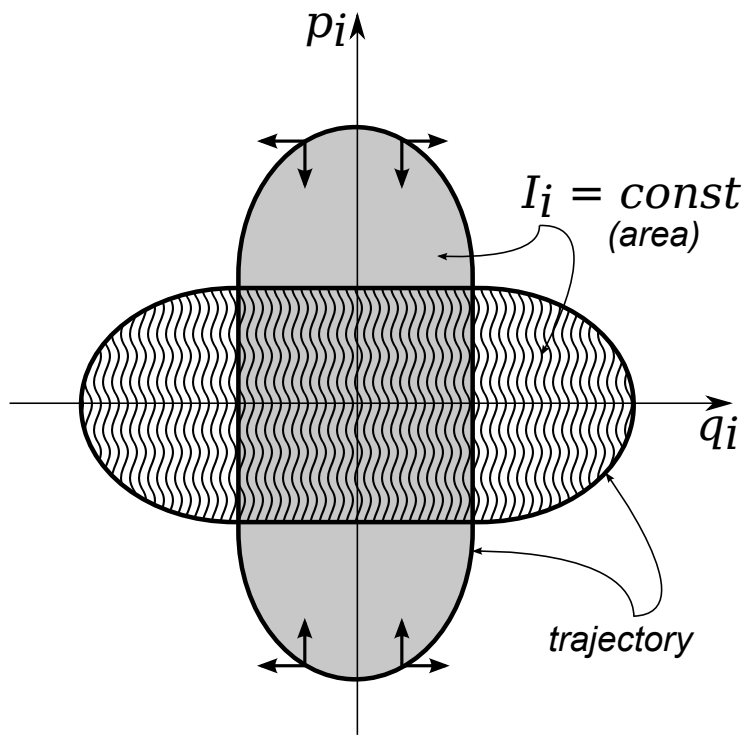}}
\caption{Schematic representation of the individual trajectory and the state space in the  beginning of the process (filled area) and in the end of the process (waved area). The abbreviated action $I_i$, which is equal to the area bounded by the trajectory, is preserved during adiabatic transformation. The arrows attached to the trajectory indicate the direction of its transformation.}\label{fig/phasespace}
\end{figure*}

We first focus on completely integrable systems. For a system in $D$-dimensional space this means that there exists $D N$ integrals of motion. The standard mechanical description uses the positions $q_i$ and the momenta $p_i \equiv m_i\,\velocity_i$ as the canonical variables, where $i$ indicates the particle number. For an integrable system it is more convenient to use the actions $I_i$ and the angles $\phi_i$ as the canonical variables. The equations of motion $\dot{q} = H_p(q,p)$ and $\dot{p} = -H_q(q,p)$ (here dot over the variable indicates the time derivative and the subscript indicates the partial derivative) upon the canonical transformation \cite{ll1} take a simple form: $\dot{I} = 0$ and $\dot{\phi} = H_{I}(I,\phi)$. This means that the action variables $I_i$ are the integrals of motion. The action variable is defined as the abbreviated action of the corresponding degree of freedom 
\begin{equation}\label{eq/02}
I_i = \frac{1}{2\pi}\,\oint\,p_i\,dq_i
\end{equation}
where the integration is taken over the trajectory of the corresponding particle during one period. Complete integrability of the system guarantees that each of the $D N$ trajectories are closed, so the integral in \eqr{eq/02} is well defined (see \figr{fig/phasespace}). 

Consider now a process, during which the volume of the dynamical system changes slowly. In this case each of the action variables introduced by \eqr{eq/02} remains constant, $\dot{I}_i = 0$, i.e. is invariant under slow volume change \cite{ll1}. Invariance of a property during slow transformation of the parameters of the dynamic system is called the adiabatic invariance and has been studied extensively for dynamical systems \cite{LochakMeunier}. The name ``adiabatic'' here shares the same meaning with the adiabatic process in thermodynamics, during which the system evolves while being isolated from the environment. 

In the following paragraphs we derive two main equations, \eqr{eq/06} and \eqr{eq/07}, for this analysis. In particular, show that the adiabatic process for the considered dynamical system is described by the following equation:
\begin{equation}\label{eq/06}
\average{K}^{\textstyle\frac{1}{2}}\,V^{\textstyle\frac{1}{D}} = J
\end{equation}
where $K \equiv \sum_i^{\sss{D N}} {m_i\,\velocity_i^2/2}$ is the total kinetic energy of the system, while $J$ depends on the abbreviated actions $I_i$ and the particles' parameters. Since every $I_i$ is constant during this process, so is $J$, which is the adiabatic invariant as well. If the transformation of the dynamical system is such that $J$ changes, the corresponding process is no longer adiabatic. However, \eqr{eq/06} still describes this process. For a general (nonadiabatic) process, during which each of $I_i$ and therefore $J$ may change, we differentiate both sides of \eqr{eq/06} and obtain
\begin{equation}\label{eq/07}
\frac{1}{2}\,\frac{d\average{K}}{\average{K}} + \frac{1}{D}\,\frac{dV}{V} = \frac{dJ}{J}
\end{equation}

We first introduce the notion of the ``trajectory average'' as the temporal average of a dynamic variable over the period of motion along a specific trajectory: $\average{x}_{\tau} \equiv \tau^{-1} \int_0^{\tau} x(t)\,dt$. We next introduce the characteristic length of the particle's trajectory $a_i \equiv \sqrt{\average{\velocity_i^2}}\,\tau_i$. The system is bounded in a box and without loss of generality we may assume that the box is cubic, so its edge length is equal to $L$ and the volume is $V = L^D$. The trajectory length is then proportional to the box length $a_i = \delta_i L$, where the proportionality coefficient $\delta_i$ is the parameter of the particular trajectory. 

Transforming the integral \eqref{eq/02} over the trajectory into the integral over the period of motion $\tau_i$ along the trajectory, we obtain $I_i = \average{K_i}_{\tau_i}\,\tau_i/\pi$, where $K_i \equiv m_i\,\velocity_i^2/2$ is the kinetic energy of the particle $i$. Multiplying both sides of this equation with $\tau_i$, we obtain that $I_i\,\tau_i = m_i\,a_i^2$. Substituting $\tau_i$ from this expression back to the expression for $I_i$ we obtain for the average kinetic energy of the particle $\average{K_i} = (2\pi^2\,I_i^2)/(m_i\,\delta_i^2\,L^2)$. 

Finally, we calculate the average kinetic energy of the entire dynamic system, $\average{\sum_i {K_i}}$. The entire dynamical system is not necessarily periodic, since the individual periods $\tau_i$ are, in general, not mutually commensurable. Because of this, the averaging time for the entire system does not equal to the period of any individual period, and may be chosen independently, representing the ``observation time'' $\tau$. For any finite $\tau$ the average $\average{\sum_i {K_i}}_{\tau}$ depends on the observation time and therefore does not reflect the properties of the dynamical system alone: $\average{\sum_i {K_i}}_{\tau} = \sum_i {\average{K_i}_{\tau_i}} + o(1/\tau)$, where $o$ is the little-o notation. In the limit of infinite time, however,  $\average{K}_{\infty} \equiv \average{\sum_i {K_i}}_{\infty} = \sum_i {\average{K_i}_{\infty}}$. Using \eqr{eq/04}, $\average{K} \equiv \average{K}_{\infty}$, so that $\average{K} = V^{-(2/D)}\,\sum_i (2\pi^2\,I_i^2)/(m_i\,\delta_i^2)$. Denoting $J^2 \equiv \sum_{i=1}^{\sss{D N}} 2\pi^2\,I_i^2/(m_i\,\delta_i^2)$ we obtain \eqr{eq/06}.

Having \eqr{eq/06} and \eqr{eq/07} established, we introduce the deterministic temperature $T_d$, the deterministic pressure $P_d$ and the deterministic entropy $S_d$ as
\begin{equation}\label{eq/08}
\begin{array}{rcl}
\displaystyle \frac{1}{2}\,T_d &\equiv& \displaystyle \frac{\average{K}}{D N}
\\\\
P_d &\equiv& \displaystyle - \left.\frac{\partial \average{H}}{\partial V}\right|_{J=const} 
\\\\
dS_d &\equiv&\displaystyle D N\,\frac{dJ}{J}
\end{array}
\end{equation}
Denoting the potential energy of the system $U \equiv H - K$, the deterministic pressure is evaluated as $P_d = N\,T_d/V - (\partial \average{U}/\partial V)_J$. Furthermore, substituting \eqr{eq/08} in \eqr{eq/07} we obtain for the average kinetic energy $d\average{K} = T_d\,dS_d - (N\,T_d/V)\,dV$ and for the average total energy (i.e. the internal energy) 
\begin{equation}\label{eq/11}
d\average{H} = T_d\,dS_d - P_d\,dV
\end{equation}
which is nothing but \eqr{eq/00}.

It is evident that the meanings of the deterministic temperature and pressure introduced by \eqr{eq/08} reflects the corresponding meaning of the thermodynamic temperature, pressure and entropy, which are revealed by statistical mechanics for stochastic systems. In particular, $1/2$ of the deterministic temperature is equal to the average kinetic energy of one degree of freedom. Note, that the notion of the deterministic temperature does not require actual equipartion of the energy, so the deterministic degrees of freedom may have different energy. Furthermore, the deterministic pressure is the sum of the ideal gas contribution and the ``configurational'' contribution. This corresponds to the exact expression for the instantaneous microscopic pressure, $P =  (3V)^{-1}\left[\sum_i\,m_i\,\velocity_i^2  + \vec{F}_i\cdot\vec{r}_i \right]$, where $\vec{F}_i$ is the total force on the particle $i$ which has the position $\vec{r}_i$. For ergodic systems this leads to the virial expression for the pressure $P = TN/V - (1/6)\,\int{r\,g(r)\,du(r)}$, where $g(r)$ is the pair correlation function and $u(r)$ is the intermolecular pair potential \cite{RowlinsonWidom}. 

%The deterministic entropy $S_d$ is a function of deterministic parameters of the system's degrees of freedom, such as the particle mass $m_i$ and the abbreviated action $I_i$. This is in contrast to the statistical description, where the entropy reflects the probabilistic nature of the statistical description and does not have a mechanical interpretation. 
%Therefore in the expression for the deterministic entropy $J$ needs to be divided by $N!$, the amount of permutations of $N$ particles. 
% $J_o$, so \eqr{eq/05} integrates to $S_d = DN\,\log(J/J_o)$. This, however, does not affect the thermodynamic behavior of the system, since $J_o$ is independent of the temperature and the volume. 

We should emphasize, that the above derivation (e.g. \eqr{eq/06} and \eqr{eq/07}) utilizes the deterministic quantities only. At no step of this derivation any thermodynamic or statistical assumption has been made. Yet, this procedure allows us to introduce the deterministic analogues of the thermodynamic variables and obtain the deterministic analogues of the thermodynamic equations. An important observation is that the entropy of a dynamical system is its adiabatic invariant.

Despite we are not using any statistical assumption, for the thermodynamic description we do not need to trace the evolution of every individual particle of the dynamical system. This means that the dynamical system allows further coarse-graining. In particular, both the Hamiltonian of the system $H$ and the adiabatic invariant $J$ are independent of order of particles counting. This means that the system has the symmetry with respect to the particles permutations or, in other words, the particles are indistinguishable, just like in the statistical description. Similarly to the classical statistical description, the deterministic entropy is defined up to an integration constant $S_o$, which is independent of the temperature and the volume. The explicit expression for the deterministic entropy, which takes into account the above arguments is therefore $S_d = S_o + \log(J^{\sss{D N}}/N!)$. Indeterminacy of $S_o$ does not affect the thermodynamic behavior of the dynamical system, since only the entropy change, but not the entropy itself, is measurable in thermodynamics. Furthermore, the factor $N!$ guarantees additivity of the entropy and reflects the indistinguishability of the particles.

\paragraph*{Illustrative example: ideal gas.}

We next illustrate our result using a specific example, the ideal gas. It is a good model system, since it represents both a mechanical system with a simple Hamiltonian allowing exact analytical calculations and a thermodynamic system which is used as a zero-approximation for many real gases. In particular, we consider $N$ point particles of mass $m$ enclosed in a $D$-dimensional cubic box of volume $V = L^D$. The particles do not interact with each other, experiencing elastic collisions with the walls only. The momentum of each particle $p_i$ remains the same between two consecutive collisions with either wall. The potential energy of the system is zero, so the total energy is $E = \sum_{i=1}^{D N}\,{p_i^2/(2m)}$. We investigate how this energy changes during the thermodynamic process of equilibrium volume variation, which for the mechanical system corresponds to slow motion of one of the box walls with a constant velocity $\welocity << \velocity_i$. 

Since there is no interaction between the particles, the motion of every degree of freedom is independent of each other. When a particle collides with the moving wall, the absolute value of its velocity changes by $2\welocity$. Every particle travels across the entire box, so the trajectory parameter $\delta_i = 2$. The distance which is passed by the moving wall between two subsequent collisions of the particle and the non-moving wall is $\Delta L = \welocity\,\left[L/(\velocity_i-\welocity) + \{L+\welocity\,L/(\velocity_i-\welocity)\}/(\velocity_i-2\welocity)\right]$. Evaluating the integral in \eqr{eq/02} over the particle's trajectory between two subsequent collisions with the non-moving wall we obtain $I_i = m\,\velocity_i\,L/\pi$. Substituting two consecutive values of $\velocity_i$ and $L$ in this expression one can show that $I_i(L) = const$, i.e. $I_i$ does not change when the wall moves, which means that $I_i$ is an adiabatic invariant. 

Evaluating the expression for the total adiabatic invariant we obtain $J = \pi\,\big[\sum_{i=1}^{D N}\,I_i^2/(2m)\big]^{1/2}$. Furthermore, a general process is described by the equation $E^{1/2}\,V^{1/D} = J$, which has the form of \eqr{eq/06} with $E$ used instead of $\average{K}$. For the process of slow variation of the system's volume (i.e. when $J=const$) the energy variation is $dE = -(2/D)\,E\,V\,dV$. For a general process (i.e. when $J$ changes), we obtain \eqr{eq/07} with $E$ used instead of $\average{K}$. Note, that for the ideal gas the equations contain instantaneous value of the total energy, not the average one. This is a consequence of the system being the ideal gas. Furthermore, for the ideal gas $\average{H} = E$. 

Introducing the deterministic temperature, the deterministic pressure and the deterministic entropy according to \eqr{eq/08}, we obtain the equation of state and the equation of the process respectively:
\begin{equation}\label{eq/16}
\begin{array}{rl}
P_d\,V =& N\,T_d
\\\\
\displaystyle P_d\,V^{1+2/D} =&  (2/D)\,J^2
\end{array}
\end{equation}
The equation of state is identical to the thermodynamic equation of state for the ideal gas. Furthermore, for $D=3$ and for the process of slow variation of the system's volume ($J=const$) the equation of process is identical to the thermodynamic adiabatic equation for the ideal gas with the adiabatic index $\gamma \equiv 1 + 2/D = 5/3$. Furthermore, it follows from the definition of $S_d$ that 
\begin{equation}\label{eq/17}
dS_{d,\, ig} = N\,\left(\frac{D}{2}\,\frac{dT_d}{T_d} + \frac{dV}{V}\right)
\end{equation}
which for $D=3$ is identical to the thermodynamic expression for the entropy variation of the ideal gas.

The dynamical system of ideal gas can be viewed as the limiting case of both the statistical description and the deterministic description. We will use this fact to establish the expression for the integration constant $S_o$ for the deterministic entropy. The statistical mechanical expression for the entropy is given in terms of the phase space volume accessible to the system \cite{Kubo}:
\begin{equation}\label{eq/18}
S_{e,\, ig} = \log \left[\frac{z_{\sss{D N}}}{N!}\frac{V^{\sss{N}}\,(2m\,E)^{\sss{D N/2}}}{(2\pi\hbar)^{\sss{D N}}}\right]
\end{equation}
where $z_{\sss{D N}} \equiv \pi^{\sss{D N/2}}/({\scriptstyle D N/2})!$ is the volume of the $D N$-dimensional unit sphere. Comparing $S_{e,\, ig}$ with $S_{d,\, ig}$ we obtain $S_o = \log z_{\sss{D N}} +  D N\,\log\left[ \sqrt{2m}/(2\pi\hbar)\right]$. 

In a general dynamical system different degrees of freedom are characterized by different values of parameters (e.g. different particle's masses), so it might be convenient to introduce new action variables $Y_i$ with the ideal gas reference. In particular, we define $Y_i \equiv (I_i/\hbar)\,(2/\delta_i)\,(m/m_i)^{1/2}$. For the ideal gas ($m_i = m$, $\delta_i = 2$) this gives $Y_i = I_i/\hbar$. The above expression for the integration constant $S_o$ represents this particular reference state, and the expression for the deterministic entropy for a general dynamical system takes the following form
\begin{equation}\label{eq/19}
S_d = \log \bigg[\frac{z_{\sss{D N}}}{N!}\bigg(\frac{1}{2^2}\sum_{i=1}^{\sss{D N}}\,Y_i^2\bigg)^{\sss{D N/2}}\,\bigg]
\end{equation}
The quantity $(\sum_{i}\,Y_i^2)^{1/2}$ can be interpreted as the diameter of the $D N$-dimensional sphere in the phase space of the dynamical system.

\paragraph*{Partially integrable systems.}

Consider now a dynamical system with $D N$ degrees of freedom which has $M$ integrals of motions, where $M < D N$. This means that the system is partially integrable and there exist only $M$ adiabatic invariants $I_i$. The Hamiltonian of this system can therefore be represented as a sum of $M$ individual terms, which correspond to $M$ integrable degrees of freedom, and the residual term which includes the remaining $D N - M$ degrees of freedom:
\begin{equation}\label{eq/21}
H_{\sss{D N}}(1, \ldots, {\scriptstyle DN}) =
\sum\limits_{i=1}^{\sss{M}} H_i(i) + H_{\sss {D N - M}}({\scriptstyle M}\!+\!1, \ldots, {\scriptstyle DN}) 
\end{equation}
If the residual term represents an ergodic system, then the above analysis can still be applied to the whole dynamical system. Indeed, a $D N - M$ dimensional ergodic system can be viewed as a single quasi-particle which performs a periodic motion in the corresponding $D N - M$ dimensional space, so the initial $DN$ dimensional dynamical system can be viewed as a new $M+1$ times integrable dynamical system.

To see this we note that the abbreviated action of a single particle $i$ defined by \eqr{eq/02} is equal to the phase space volume of that particle $\Omega_i \equiv (2\pi)^{-1}\,\int\int\,dp_i\,dq_i$, since $\oint\,p_i\,dq_i = \int\int\,dp_i\,dq_i$ (see \figr{fig/phasespace}). The motion of this particle along the phase space surface, which bounds the corresponding phase space volume, is ergodic. This means that the phase space volume $\Omega_i$ available to a single particle is an adiabatic invariant. For an ergodic system of $G$ particles the phase space volume $\Omega_{\sss{G}}(1,\ldots,{\scriptstyle G}) \equiv (2\pi)^{-\sss{G}}\,\prod_{g=1}^{\sss{G}}\int\int\,dp_g\,dq_g$ is an adiabatic invariant as well \cite{Hertz1910a, Hertz1910b} and can therefore be represented by a single variable $\Omega_{\sss{G}}$ which is the analogue of the abbreviated action for an ergodic system. 

For the partially integrable system we therefore can identify $M$ abbreviated actions $I_i$ which correspond to the integrable degrees of freedom, and one abbreviated action which corresponds to the ergodic system described by the residual Hamiltonian $H_{\scriptscriptstyle D N - M}$, having in total $M+1$ abbreviated actions. The initial partially integrable system with $D N$ degrees of freedom can therefore be reduced to a completely integrable system with $M+1$ degrees of freedom corresponding to $M+1$ quasi-particles. The analysis of integrable dynamical systems is then applied directly for the new system, replacing $D N$ in all the formulas with $M+1$. In the limiting case of completely ergodic system we have $M=0$, and all the expressions in this paper reduce to the standard statistical mechanical expressions.

Note, that the deterministic temperature and pressure of a partially integrable system defined by \eqr{eq/08} do not necessarily coincide with the statistical temperature and pressure of the corresponding ergodic subsystem. Such state, being formally a non-equilibrium one, is, however, stable: because the integrable degrees of freedom are decoupled from the ergodic subsystem, there is no means for the system to ``equilibrate'', i.e. to equalize the corresponding temperatures and the pressures. Still, the evolution of the whole system is described by \eqr{eq/00}, the statement of the first and the second laws of thermodynamics for equilibrium processes, which means that the  system as a whole is nevertheless in equilibrium. A notable example of such system is nonthermal plasma \cite{Kerrebrock1964}.

\paragraph*{Concluding remarks.}

The above analysis shows that an arbitrary dynamical system which is described by a classical Hamiltonian can be thermalized, i.e. reveals the thermodynamic behavior. While stochasticity of many deterministic dynamical systems is realized via dynamical chaos, so they are naturally thermalized, we show that non-stochastic dynamical systems undergo thermalization as well. This is especially important as it allows one to not investigate thermalization of each particular system. Rather, we have proved a general result for any classical Hamiltonian system. 

It is important to realize that in order the system the be able to thermalize, its underlying dynamics needs to be Hamiltonian. In particular, it should be described by the Hamiltonian equations of motion, while the Hamiltonian itself has to consist of the kinetic component, which is second order in the particles momenta, and the potential component, which depends on the particles positions only. In contrast, no stochasticity is required. Furthermore, the thermodynamic limit is not required: neither the volume nor the number of particles need to be large in order for the system to undergo thermalization. In fact, the number of particles can be as small as one, and we are still able to define the temperature and the entropy in the same way, and also able to derive the fundamental thermodynamic relation \eqref{eq/00}. This also emphasizes the role of the so-called observation time $\tau$. The exact results are formulated for the ``macroscopic'' variables, which are the averages of the corresponding ``microscopic'' variables over infinite time. For any finite $\tau$ the accuracy of this averaging increases with $\tau$, and when $\tau$ is much larger than the characteristic microscopic time, it is safe to impose that $\tau = \infty$.  This is the case, in particular, for slow transformations of the system, and is similar to problem of the observation time in statistical mechanics.

Finally, we emphasize the role of the adiabatic invariance. Classical equilibrium thermodynamics is valid for slow and gradual transformations of the system, which are exactly the conditions for adiabatic transformations of a dynamical system. Furthermore, our analysis utilizes the direct correspondence between a thermodynamic process, during which the system is thermodynamically isolated from the environment, and a mechanical process, during which the system possesses one (or many) mechanical adiabatic invariants. In particular, we have demonstrated that the entropy of a thermodynamic system is the adiabatic invariant of the corresponding dynamical system.

V.K. is grateful to Mr.~Konstantin Yun for financial support of the research.

\bibliographystyle{unsrt}

%\bibliography{adiabat}
%\input{variational.bbl}

\end{document}